%% file: higherOrderSC_CDS.tex
\documentclass[ALICE,manyauthors]{cernphprep}

\usepackage[comma,square,numbers,sort&compress]{natbib}
\usepackage{hyperref}
\usepackage{lineno}
\usepackage[T1]{fontenc}

\usepackage{color}
\definecolor{dgreen}{cmyk}{1.,0.,1.,0.4} % dark green
\definecolor{orange}{cmyk}{0.,0.353,1.,0.} % orange

\begin{document}%

%\linenumbers % TBI comment out eventually 

%%%%%%%%%%%%%%%  Title page %%%%%%%%%%%%%%%%%%%%%%%%
\begin{titlepage}
\PHyear{2021}
\PHnumber{005}      % required, will be obtained from PH
\PHdate{6 January}  % required, will be obtained from PH
%

%%% Put your own title + short title here:
\title{Multi-harmonic correlations of different flow amplitudes in Pb--Pb collisions at $\sqrt{s_{_{\rm NN}}}=2.76$~TeV}
\ShortTitle{Multi-harmonic correlations of different flow amplitudes...} % appears on right page headers

%%% Do not change the next lines
\Collaboration{ALICE Collaboration\thanks{See Appendix~\ref{app:collab} for the list of collaboration members}}
\ShortAuthor{ALICE Collaboration} % appears on left page headers, do not change

\begin{abstract}
\input{Abstract.tex}
\end{abstract}
\end{titlepage}
\setcounter{page}{2}

\input{Section_1.tex}

\input{Section_2.tex}

\input{Section_3.tex}

\input{Section_4.tex}

\input{Section_5.tex}

%================================================================================

%%%%% acknowledgements
\newenvironment{acknowledgement}{\relax}{\relax}
\begin{acknowledgement}
\section*{Acknowledgements}
\noindent The ALICE Collaboration would like to thank Harri Niemi for providing the latest predictions from the state-of-the-art hydrodynamic model.
%{\it (TBI: The default acknowledgments will be done by EB centrally.)}
\input{fa_2020-12-27.tex}    %%%%%%% done by webmaster team
\end{acknowledgement}

%%%%%%%% Bibliography (In case of using bibtex generate the bbl requested by arXiv)
\bibliographystyle{utphys} % Remember we use title in the biblio
\nocite{*}
\bibliography{bibliography}
%\input {bibliography.tex}  

%%%%%%%%% appendix with author list
\newpage
\appendix
\section{The ALICE Collaboration}
\label{app:collab}
\input{Alice_Authorlist_2020-12-27.tex}  %%%%%%% done by webmaster team
\end{document}

%% file: Abstract.tex
The event-by-event correlations between three flow amplitudes are measured for the first time in Pb--Pb collisions, using higher-order Symmetric Cumulants.
We find that different three-harmonic correlations develop during the collective evolution of the medium, when compared with correlations that exist in the initial state. These new results cannot be interpreted in terms of previous lower-order flow measurements, since contributions from two-harmonic correlations are explicitly removed in the new observables. Comparison with Monte Carlo simulations provides new and independent constraints for the initial conditions and system properties of nuclear matter created in heavy-ion collisions.

%% file: Section_1.tex
Under conditions of extreme temperature and density, the fundamental theory of the strong interaction, quantum chromodynamics (QCD), predicts the existence of a quark--gluon plasma (QGP). In this state, quarks are deconfined from hadrons, but contrary to the initial theoretical expectations, remain strongly coupled and form a liquid state~\cite{Jacak:2010zz}.
Results from heavy-ion collision data are consistent with the scenario in which the produced nuclear matter undergoes collective expansion, dominated by its hydrodynamic response to the anisotropies in the initial state geometry.
This phenomenon is known as anisotropic flow~\cite{Ollitrault:1992bk}. This collective dynamics is sensitive to $\eta/s$ and $\zeta/s$, where $\eta$ and $\zeta$ are shear and bulk viscosities, and $s$ the entropy density.
The successful description of heavy-ion data with hydrodynamic models was essential to determine the low value of $\eta/s$ of the QGP~\cite{Kovtun:2004de} and established the perfect liquid paradigm, one of the most striking recent discoveries in high-energy physics~\cite{Heinz:2013th,Braun-Munzinger:2015hba,Busza:2018rrf}.

In models that describe heavy-ion collisions, the produced matter evolves collectively, with particles being emitted independently along the azimuthal direction with a distribution $f(\varphi)$. The corresponding Fourier series is given by
\begin{equation}
f(\varphi)=\frac{1}{2\pi}\left[1+2\sum_{n=1}^{\infty}v_n\cos[n(\varphi-\Psi_n)]\right],
\end{equation}
where the flow amplitude $v_n$ and the symmetry plane angle $\Psi_n$ designate the magnitude and orientation of the $n$th order anisotropic flow~\cite{Voloshin:1994mz}.
Experimental challenges of measuring these anisotropic flow observables are overcome with the development of multiparticle azimuthal correlations~\cite{Wang:1991qh,Jiang:1992bw,Borghini:2001vi,Bilandzic:2010jr,Bilandzic:2013kga}.
A great deal of additional information can be extracted from correlations between different flow amplitudes and/or different symmetry planes~\cite{Niemi:2012aj,Aad:2014fla,Jia:2014jca,Aad:2015lwa,Qian:2016pau}.

The correlations between event-by-event fluctuations of two different flow amplitudes were quantified with the \textit{Symmetric Cumulants}~(SC) observables~\cite{Bilandzic:2013kga,ALICE:2016kpq}, defined by SC$(k,l) \equiv \langle v_k^2 v_l^2\rangle - \langle v_k^2\rangle\langle v_l^2\rangle$, with the angular brackets denoting an average over all events. 
The measurements of their centrality and transverse momentum  ($p_{\rm T}$) dependences revealed that correlations among different flow magnitudes depend on harmonic orders as well as the collision centrality, while showing moderate $p_{\rm T}$ dependence in semicentral collisions.
The results in Refs.~\cite{Bilandzic:2013kga,ALICE:2016kpq} showed that the different SC($k,l$) observables have different sensitivities to the initial conditions of a heavy-ion collision and properties of the created system, and can therefore help in separating the effects of $\eta/s$ in the final state anisotropies from the contributions originating in the initial state. Furthermore, it was demonstrated that the SC observables are more sensitive to the temperature dependence $\eta/s(T)$ than the individual flow amplitudes, which are sensitive only to the average values $\left<\eta/s\right>$~\cite{Niemi:2015qia,ALICE:2016kpq}.

%% file: Section_2.tex
In this paper, a new set of observables, dubbed {\it higher order SC}, are analyzed~\cite{Mordasini:2019hut}. These higher order observables extract the genuine correlation among multiple flow amplitudes, and provide new and independent constraints for both the initial conditions and the QGP properties.
The genuine correlation (or cumulant) of three flow amplitudes, where lower-order two-harmonic correlations have been removed, can be obtained with the following expression~\cite{Kubo:1962,Mordasini:2019hut}:
\begin{equation}
\mathrm{SC}(k,l,m) \equiv \left<v_k^2v_l^2v_m^2\right> 
- \left<v_k^2v_l^2\right>\left<v_m^2\right>
- \left<v_k^2v_m^2\right>\left<v_l^2\right>
- \left<v_l^2v_m^2\right>\left<v_k^2\right>
+ 2 \left<v_k^2\right>\left<v_l^2\right>\left<v_m^2\right>\,.
\label{eq:SC-three-amplitudes-th}
\end{equation}
The observable SC($k,l,m$) is, by definition, the 3rd order cumulant of three flow amplitudes $v_k^2$, $v_l^2$ and $v_m^2$. If the previously used low order flow observables, like $v_n\{2\}$, $v_n\{4\}$~\cite{Borghini:2001vi} or SC($k,l$)~\cite{Bilandzic:2013kga}, would be able to characterize all collective correlations and anisotropic flow in the system, SC($k,l,m$) would be identically zero.
On the contrary, the non-vanishing results for SC($k,l,m$) provide access to the information to which these traditionally used flow observables are insensitive. The normalized versions of these observables are defined as
\begin{equation}
\text{NSC}(k,l,m)\equiv\frac{\text{SC}(k,l,m)}{\langle v_k^2 \rangle \langle v_l^2 \rangle \langle v_m^2\rangle}\,,
\label{eq:NSC-definition}
\end{equation}
which makes it easier to identify the origin of the correlations, either from the initial stage or from the collective expansion~\cite{Mordasini:2019hut}.

Another important aspect is the sign of the SC($k,l,m$) observables which is not trivial and can be understood if the definition in Eq.~(\ref{eq:SC-three-amplitudes-th}) is rewritten as:
\begin{equation}
\mathrm{SC}(k,l,m) = \left<\left(v_k^2 - \left<v_k^2\right>\right)\left(v_l^2 - \left<v_l^2\right>\right)\left(v_m^2 - \left<v_m^2\right>\right)\right>\,.
\label{eq:SC-three-amplitudes-th-refefinition}
\end{equation}
For SC$(k,l,m) > 0$ there are the following two distinct possibilities: a) if in an event it was found that $v_k^2 > \left<v_k^2\right>$ and $v_l^2 > \left<v_l^2\right>$, then the probability to find $v_m^2 > \left<v_m^2\right>$ in that event is enhanced (this case is marked as $(+,+,+)$ pattern in the event-by-event flow fluctuations); b) if $v_k^2 > \left<v_k^2\right>$ and $v_l^2 < \left<v_l^2\right>$ in an event, that enhances the probability to find $v_m^2 < \left<v_m^2\right>$ in that event and this is marked as $(+,-,-)$ pattern.
By using the same reasoning, it can be concluded that SC$(k,l,m) < 0$ permits only the $(+,+,-)$ and $(-,-,-)$ patterns.
These persistent patterns of event-by-event flow fluctuations are invariant with respect to permutations of amplitudes of flow harmonics in the definition of SC$(k,l,m)$, and they are a direct imprint of the three-harmonic correlations. 

It was demonstrated in Ref.~\cite{Mordasini:2019hut} that SC$(k,l,m)$, as defined in Eq.~(\ref{eq:SC-three-amplitudes-th}), can be estimated reliably in an experiment with the following combination of azimuthal correlators:
\begin{eqnarray}
\mathrm{SC}(k,l,m) &=&\left<\left<\cos[k\varphi_1\!+\!l\varphi_2\!+\!m\varphi_3\!-\!k\varphi_4\!-\!l\varphi_5\!-\!m\varphi_6]\right>\right>\nonumber\\
&-&\left<\left<\cos[k\varphi_1\!+\!l\varphi_2\!-\!k\varphi_3\!-\!l\varphi_4]\right>\right>\left<\left<\cos[m(\varphi_5\!-\!\varphi_6)]\right>\right>\nonumber\\
&-&\left<\left<\cos[k\varphi_1\!+\!m\varphi_2\!-\!k\varphi_5\!-\!m\varphi_6]\right>\right>\left<\left<\cos[l(\varphi_3\!-\!\varphi_4)]\right>\right>\nonumber\\
&-&\left<\left<\cos[l\varphi_3\!+\!m\varphi_4\!-\!l\varphi_5\!-\!m\varphi_6]\right>\right>\left<\left<\cos[k(\varphi_1\!-\!\varphi_2)]\right>\right>\nonumber\\
&+&2\left<\left<\cos[k(\varphi_1\!-\!\varphi_2)]\right>\right>\left<\left<\cos[l(\varphi_3\!-\!\varphi_4)]\right>\right>\left<\left<\cos[m(\varphi_5\!-\!\varphi_6)]\right>\right>\,.
\label{eq:SC-three-amplitudes-exp}
\end{eqnarray}
The double average notation indicates that in the first step averaging is performed over all distinct combinations of 2, 4, or 6 particles within the same event, and then these results are averaged over all events. Each azimuthal correlator in the above estimator can be measured efficiently and exactly with the Generic Framework published in Ref.~\cite{Bilandzic:2013kga}. By definition, this estimator ensures that large systematic biases from self-correlations and symmetry planes $\Psi_n$ are eliminated. In the absence of nonflow (correlations between a few particles unrelated to collective phenomena and anisotropic flow), it reduces analytically to Eq.~(\ref{eq:SC-three-amplitudes-th}), even for the case of large event-by-event flow fluctuations~\cite{Mordasini:2019hut}.

%% file: Section_3.tex
The results presented in this paper are obtained with the data from Pb--Pb collisions at $\sqrt{s_{\mathrm {NN}}} = 2.76$~TeV collected with the ALICE detector in 2010. After the event and track selection, the data sample corresponds to about $8.2~\times~10^{6}$~minimum bias events for the $0$--$50\%$ centrality range. The Pb--Pb dataset from 2011 is not included due to the significantly different detector and trigger conditions.

Detailed descriptions of the ALICE detector and performance can be found in Refs.~\cite{Aamodt:2008zz,Carminati:2004fp,Alessandro:2006yt,Abelev:2014ffa}.
The time projection chamber (TPC) was used to reconstruct charged particles and measure their momenta~\cite{Alme:2010ke}. The inner tracking system (ITS) was used to improve the vertex determination and momentum resolution, while its innermost part, the silicon pixel detector (SPD)~\cite{Dellacasa:1999kf,Aamodt:2010aa}, provided the default centrality estimation.
Two scintillator arrays (V0A and V0C) were used for triggering and for an alternative determination of centrality~\cite{Aamodt:2010cz,Abbas:2013taa,Adam:2015ptt}. 
The trigger conditions are identical to those described in Refs.~\cite{Aamodt:2010pa, Aamodt:2010cz}.

The event and track selection are based on previous SC analyses~\cite{ALICE:2016kpq, Acharya:2017gsw}. The reconstructed primary vertex (PV) is required to be within $\pm 10$~cm from the nominal interaction point along the beam axis.
The main analysis is performed using tracks reconstructed only with the TPC (referred to as TPC-only from now on) in the kinematic range $0.2 < p_{\mathrm T} < 5.0$~GeV/$c$ and $\vert \eta \vert < 0.8$. The low $p_{\mathrm T}$ cutoff decreases the biases from the smaller reconstruction efficiency, while the high $p_{\mathrm T}$ cutoff reduces the anisotropic contaminations in the azimuthal distributions from jets.
The selected tracks are reconstructed with a minimum of 70 space points out of a maximum of 159 in TPC and the $\chi^2/NDF$ of their momentum fit is required to satisfy $0.1 < \chi ^2/NDF < 4.0$.
Only tracks with a maximum distance of closest approach (DCA) to the primary vertex of $2.4$~cm in the transverse plane and $3.2$~cm along the beam axis are kept for the analysis. This choice  reduces the contributions from secondary tracks and has already been used in Ref.~\cite{ALICE:2016kpq} with hybrid tracks, for which the tracking information is combined from both the TPC and the ITS detectors to achieve the best transverse momentum resolution and to correct for the non-uniform azimuthal acceptance due to dead zones in the SPD~\cite{Abelev:2014ffa,Acharya:2020rlz}.
Also, tracks with abrupt change of direction, e.g. due to multiple scattering or $K^\pm$ decays, are rejected. With this selection, the contamination from secondaries in TPC-only tracks varies from about $16$\% at $0.2$~GeV/$c$ to about $7$\% at $5$~GeV/$c$. The track reconstruction efficiency is almost constant at about 80--88\% as a function of transverse momentum. Its uncertainties are found to be negligible and thus not propagated in the final results.

Corrections both for non-uniform reconstruction efficiency (NUE) as a function of transverse momentum and non-uniform acceptance (NUA) as a function of azimuthal angle are computed as particle weights, following Ref.~\cite{Bilandzic:2013kga}. Particle weights for NUE were obtained with the Monte Carlo generator HIJING~(Heavy-Ion Jet INteraction Generator)~\cite{Gyulassy:1994ew}, while the ones for NUA are data driven.
Only the corrections for NUE are applied to all the selected tracks in the main analysis with the default selection. Effects of NUA in TPC-only tracks were also checked, but found to be neligible.
The nonflow contributions estimated with HIJING are found to be negligible for all SC($k,l,m$) observables reported in this paper~\cite{Mordasini:2019hut}.

The systematic uncertainties are estimated by varying each selection criterion independently.
The values of ${\rm SC}(k,l,m)$ with the variation and with the default selection are compared in each centrality interval. If the difference between the two results when taking into account the correlations between their statistical uncertainties is larger than one $\sigma$ ($\sigma$ is the uncertainty of the difference), the variation is included in the quadratic sum for the total systematic uncertainty.
The importance of each trial depends on the considered SC($k,l,m$).
The data sample was collected with two configurations of the magnetic field polarity in the solenoid magnet surrounding the ALICE central barrel detectors, giving two samples with similar size. 
The main analysis uses both samples, and no significant systematic effect is seen for the analysis on each individual orientation of the field polarity.
Below, the ranges of relative variations observed in semicentral collisions (20--50\%) for each trial are reported.
Moreover, the variations observed in collisions with a centrality up to 20\%, and for ${\rm SC}(2,4,6)$ and ${\rm SC}(3,4,5)$ in the range 20--30\%, can be larger than the ones indicated due to the small size of the signal and are therefore not reported. The systematic uncertainties are represented by the shaded boxes around each data point in all figures.

On the other hand, there are variations which impact only some SC($k,l,m$) observables. For example, the variation of the distance of the PV to the nominal interaction point along the beam direction ($\pm 6$~cm and $\pm 12$~cm) does not impact ${\rm SC}(2,3,5)$, ${\rm NSC}(2,3,5)$ and ${\rm SC}(3,4,5)$, but results to an uncertainty of about $3.2$\% for ${\rm SC}(2,3,4)$ and ${\rm NSC}(2,3,4)$.
For the DCA variation in the plane transverse to the beam direction (from $2.4$~cm to $1$~cm and $2$~cm) only ${\rm SC}(2,4,6)$ is not affected, while there is an effect of about $12$\% for ${\rm NSC}(2,3,4)$ to about $36$\% for ${\rm SC}(2,3,5)$.
The default analysis uses the centrality estimated with the SPD, while the systematic check is based on the determination of the centrality with the V0 detector. This change impacts the final results for all combinations with the exception of ${\rm SC}(3,4,5)$, ranging from about $15$\% for ${\rm SC}(2,3,4)$ and ${\rm NSC}(2,3,4)$ to $21$\% for ${\rm SC}(2,3,5)$.
The variation of the minimum number of space points in the TPC (from 70 to 50 and 100 space points) leads to systematic biases in ${\rm SC}(2,3,4)$, ${\rm SC}(2,3,5)$ and ${\rm NSC}(2,3,5)$, ranging from $5$\% for ${\rm SC}(2,3,4)$ to $14$\% for ${\rm SC}(2,3,5)$.
This is also the case for the quality of fit $\chi^2/NDF$ for $0.3 < \chi^2/NDF < 4.0$ and $0.1 < \chi^2/NDF < 3.5$. This leads to significant differences for ${\rm SC}(2,4,6)$, ${\rm SC}(3,4,5)$ and ${\rm NSC}(2,3,5)$ (about $12$\% for ${\rm NSC}(2,3,5)$). For the tightening of the DCA criterion along the beam axis from $3.2$~cm to $2.1$~cm, we report the systematic bias of about $8$--$10$\% for ${\rm SC}(2,3,5)$ and ${\rm NSC}(2,3,5)$.
Finally, non-negligible systematic effects are seen when repeating the analysis with hybrid tracks, which have a smaller contamination from secondaries, allowing an estimation of their systematic effects in the default selection. For this last check, all SC($k,l,m$) see significant changes (between $4$\% and $19$\% for ${\rm SC}(2,3,4)$ and ${\rm NSC}(2,3,5)$, respectively).

%% file: Section_4.tex
\begin{figure}[h!]
	\centering
	\includegraphics[width=0.7\linewidth]{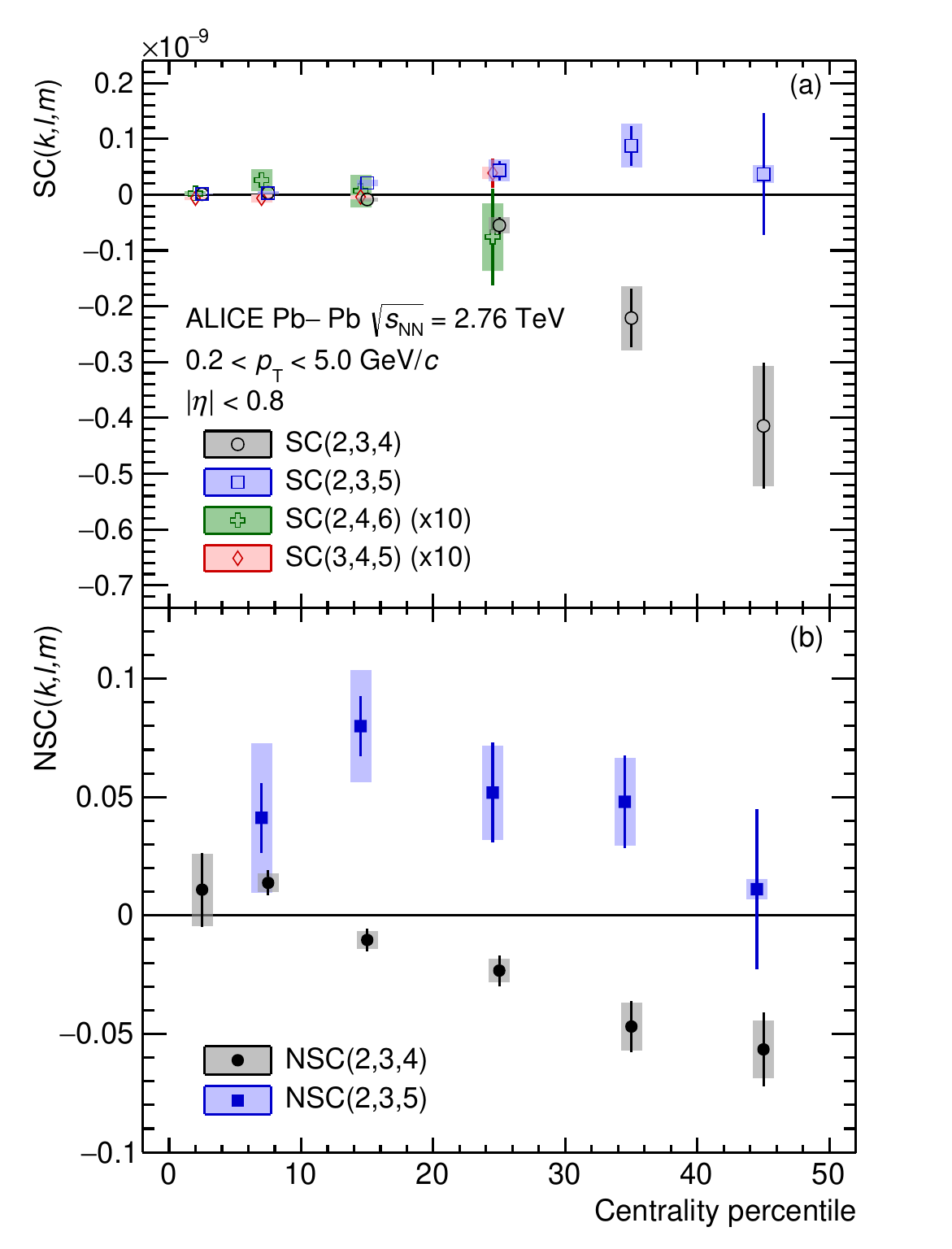}
	\caption{Centrality dependence of SC(2,3,4), SC(2,3,5), SC(2,4,6) and SC(3,4,5) (a) and of NSC(2,3,4) and NSC(2,3,5) (b) in Pb--Pb collisions at $\sqrt{s_{_{\rm NN}}}=2.76$~TeV. The statistical (systematic) uncertainties are shown with the lines (boxes). The data points are shifted horizontally for visibility.}
	\label{s:Results_fig1}
\end{figure}

The centrality dependence of SC($k,l,m$) and NSC($k,l,m$) for the different combinations of flow amplitudes is shown in Fig.~\ref{s:Results_fig1} (a) and Fig.~\ref{s:Results_fig1} (b), respectively.
When moving from central to semicentral collisions, the magnitude of both ${\rm SC}(2,3,4)$ and ${\rm SC}(2,3,5)$ increases, albeit with opposite sign.
These non-zero values for semicentral collisions are the first experimental indications of correlations between three flow amplitudes.
The results for ${\rm SC}(2,3,5)$ provide new and independent constraints on the non-linear response contribution in $v_5$ from $v_2$ and $v_3$, which for the first time do not require any assumption in the derivation on the nature of two-harmonic correlations~\cite{Yan:2015jma}.
For the higher order flow amplitudes, the measurements for ${\rm SC}(2,4,6)$ and ${\rm SC}(3,4,5)$ are compatible with zero for all centralities.
The negative increasing trend observed for ${\rm SC}(2,3,4)$ is also present for ${\rm NSC}(2,3,4)$. However, this is not the case for ${\rm SC}(2,3,5)$ and ${\rm NSC}(2,3,5)$. The increase seen in the former cannot be found in the latter, which shows a decrease for semicentral events. This different behavior originates from the fact that the non-linear response introduces a correlation among all three amplitudes in SC(2,3,5), while the contribution from non-linear response is not present in SC(2,3,4).

\begin{figure}[h!]
	\centering
	\includegraphics[width=0.9\linewidth]{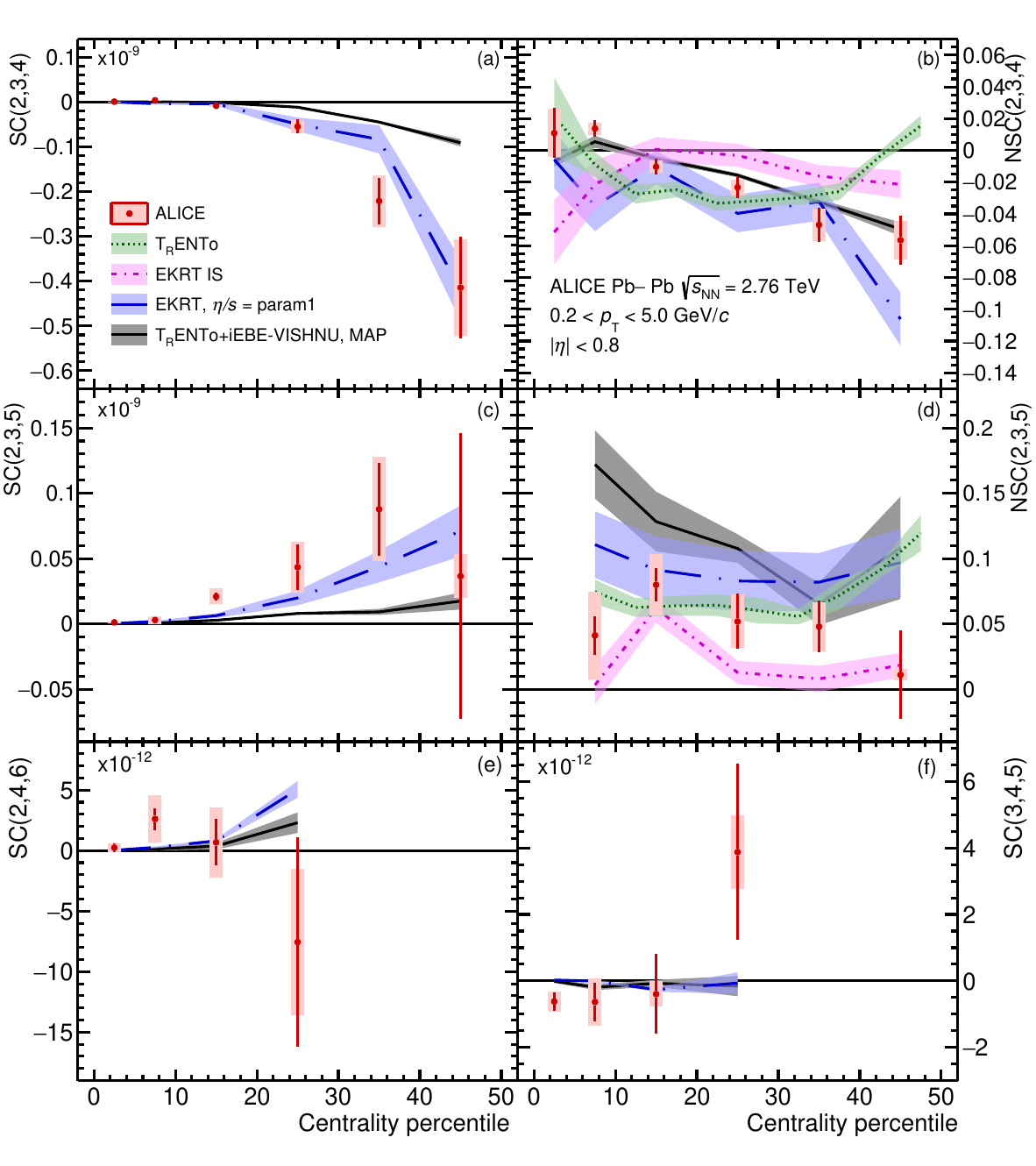}
	\caption{Predictions from the hydrodynamical models for the centrality dependence for the SC($k,l,m$) [panels (a), (c), (e) and (f)] and NSC($k,l,m$) [panels (b) and (d)] in Pb--Pb collisions at $\sqrt{s_{_{\rm NN}}}=2.76$~TeV. The statistical uncertainties are shown with coloured bands. The predictions are compared with the ALICE results from Fig.~\ref{s:Results_fig1} shown with red markers. The bands represent the statistical uncertainty of each model.}
	\label{s:Results_fig2}
\end{figure}

The results for the higher order SC observables are compared with the event-by-event Eskola-Kajantie-Ruuskanen-Tuominen~(EKRT)+viscous~\cite{Niemi:2015qia} and T$_{\rm{R}}$ENTo + iEBE-VISHNU hydrodynamic models~\cite{Bernhard:2019bmu}.
In the EKRT model, the initial energy density profiles are calculated using a next-to-leading order perturbative-QCD+saturation model~\cite{Paatelainen:2012at,Paatelainen:2013eea}. The subsequent space--time evolution is described by relativistic dissipative fluid dynamics with different temperature parameterizations $\eta/s(T)$. 
This state-of-the-art model gives a good description of the charged hadron multiplicity and the low-$p_{\rm T}$ region of the charged hadron spectra at BNL's Relativistic Heavy Ion Collider and at CERN's Large Hadron Collider.
Each of the $\eta/s(T)$ parameterizations is adjusted to reproduce the measured $v_n$ from central to semiperipheral collisions.
The model calculations in which the temperature of the phase transition is larger than for ``param1'' parameterization are ruled out by the previous measurements~\cite{ALICE:2016kpq,Acharya:2017gsw}. In the study presented in this paper, the EKRT prediction for the centrality dependence of SC($k,l,m$) was obtained from a sample consisting of 40k events in the 0--100\% centrality range.

The calculations for the $\eta/s(T)=\mathrm{``param1}$'' parametrisation, which gives a good description of the lower order SC results, are thus compared to our new results for higher order SC in Fig.~\ref{s:Results_fig2}. They can describe the overall trends of all combinations in the centrality dependence. However, ${\rm SC}(2,4,6)$ is found to be strictly positive in models.

The hybrid hydrodynamic model T$_{\rm{R}}$ENTo+iEBE-VISHNU has successfully described the previous ALICE measurements~\cite{Bernhard:2019bmu}. It consists of the T$_{\rm{R}}$ENTo model~\cite{Moreland:2014oya} for the initial condition, which is connected with a free streaming to a 2+1 dimensional causal hydrodynamic model VISH2+1~\cite{Shen:2014vra,Song:2007ux}. The evolution is continued in the hadronic phase via the UrQMD model~\cite{Bass:1998ca,Bleicher:1999xi}. The  initial conditions, $\eta/s(T)$, $\zeta/s(T)$ and other free parameters of the hybrid model are extracted by the global Bayesian analysis. We perform a model calculation with the best-fit parameter points chosen by maximum a posteriori (MAP) for Pb--Pb collisions at $\sqrt{s_{\text{NN}}}=2.76$~TeV as they are reported in Ref.~\cite{Bernhard:2019bmu}. All the kinematic cuts such as transverse momentum and pseudorapidity intervals are matched with the data reported in this paper.

In heavy-ion collisions, the main source of anisotropy in the azimuthal distribution in the final state originates from anisotropies in the initial state geometry. The initial state geometry can be described by quantities called eccentricities $\epsilon_n$ which are the moments of the initial energy (or entropy) density. For instance, the values of $\epsilon_2$ and $\epsilon_3$ indicate to what extent the initial geometry is elliptical and triangular, respectively. For small values of eccentricities, one can approximate the response of the collective evolution to the initial state geometry as a linear relation $v_n = k_n\epsilon_n$~\cite{Teaney:2010vd,Gardim:2011xv}. For $n=2,3$, this linear approximation is more accurate than for higher harmonics where non-linear terms play a non-negligible role~\cite{Niemi:2012aj}. If the higher order eccentricity cumulants are normalized by their averages (analogous to Eq.~(\ref{eq:NSC-definition})), the response coefficients $k_n$ can cancel between numerator and denominator. Therefore, any difference in the NSC values calculated from the eccentricities in the initial state to those obtained from the measured flow amplitudes in the final state is an indication of a hydrodynamic non-linear response. 

The comparison to the T$_{\rm{R}}$ENTo+iEBE-VISHNU calculation is also shown in Fig.~\ref{s:Results_fig2}.
The overall trends in the centrality dependence are captured by this model. However, both SC(2,3,4) and SC(2,3,5) are clearly underestimated, while NSC(2,3,4) and NSC(2,3,5) are in a better agreement with the data.
In the case of NSC($k$,$l$,$m$), predictions from T$_{\rm{R}}$ENTo for the initial state are shown in Fig.~\ref{s:Results_fig2}~(b) and Fig.~\ref{s:Results_fig2}~(d). As iEBE-VISHNU uses T$_{\rm{R}}$ENTo as input, the comparisons between the two sets of predictions can give insights about the development of multi-harmonic correlations in the system.
The relative change in NSC(2,3,4) for iEBE-VISHNU calculations from the ones from T$_{\rm{R}}$ENTo for 10--30\% centralities indicates that in addition different correlations have developed during the hydrodynamic evolution of the medium. 
The same phenomenon is hinted at within uncertainties in NSC(2,3,5). In this latter case, this can be explained by the non-linear response contribution to $v_5$ induced by the low order $v_2$ and $v_3$ found in Refs.~\cite{Acharya:2017zfg,Acharya:2020taj}.
For SC(2,4,6) and SC(3,4,5), iEBE-VISHNU is in agreement with the predictions from EKRT within uncertainties.

Recent Bayesian analyses~\cite{Bernhard:2016tnd, Bernhard:2019bmu} show that the T$_{\rm{R}}$ENTo model reproduces certain features of EKRT models with the energy deposition parameter, p $\approx$ 0.0. %\pm$ 0.1.
However, as it is shown in Fig.~\ref{s:Results_fig2}~(b) and Fig.~\ref{s:Results_fig2}~(d), in semicentral collisions the T$_{\rm{R}}$ENTo model shows stronger initial-state correlations among eccentricities than the EKRT model,  and the resulting final-state multi-harmonic correlations obtained with SC($k$,$l$,$m$) show differences as well.
This difference can originate from the fact that EKRT does not include effects from bulk viscosity while the extracted bulk viscosities from two different Bayesian analyses give sizable differences.

%% file: Section_5.tex
In summary, we have presented the first measurements of correlations between three flow amplitudes, obtained with higher order SC observables in Pb--Pb collisions at $\sqrt{s_{\rm {NN}}} = 2.76$~TeV.
The non-zero values of SC$(k,l,m)$ for semicentral collisions are the first experimental indication of correlations (cumulants) between three flow amplitudes. 
The relative changes between T$_{\rm{R}}$ENTo and iEBE-VISHNU for NSC(2,3,4) and NSC(2,3,5) are consistent with the development of different correlations during the collective evolution of the medium.
A similar conclusion can be extracted from the EKRT model.
These results provide the first constraints on the non-linear response contribution in $v_5$ from $v_2$ and $v_3$, which do not require any assumption on the nature of lower-order two-harmonic correlations.
The new results for SC$(k,l,m)$ provide independent constraints for the initial conditions, system properties, non-linear response and possible patterns of event-by-event flow fluctuations, when compared to the previous flow measurements obtained with lower-order observables.

%% file: fa_2020-12-27.tex
% Version: 2020-12-27

The ALICE Collaboration would like to thank all its engineers and technicians for their invaluable contributions to the construction of the experiment and the CERN accelerator teams for the outstanding performance of the LHC complex.
The ALICE Collaboration gratefully acknowledges the resources and support provided by all Grid centres and the Worldwide LHC Computing Grid (WLCG) collaboration.
The ALICE Collaboration acknowledges the following funding agencies for their support in building and running the ALICE detector:
A. I. Alikhanyan National Science Laboratory (Yerevan Physics Institute) Foundation (ANSL), State Committee of Science and World Federation of Scientists (WFS), Armenia;
Austrian Academy of Sciences, Austrian Science Fund (FWF): [M 2467-N36] and Nationalstiftung f\"{u}r Forschung, Technologie und Entwicklung, Austria;
Ministry of Communications and High Technologies, National Nuclear Research Center, Azerbaijan;
Conselho Nacional de Desenvolvimento Cient\'{\i}fico e Tecnol\'{o}gico (CNPq), Financiadora de Estudos e Projetos (Finep), Funda\c{c}\~{a}o de Amparo \`{a} Pesquisa do Estado de S\~{a}o Paulo (FAPESP) and Universidade Federal do Rio Grande do Sul (UFRGS), Brazil;
Ministry of Education of China (MOEC) , Ministry of Science \& Technology of China (MSTC) and National Natural Science Foundation of China (NSFC), China;
Ministry of Science and Education and Croatian Science Foundation, Croatia;
Centro de Aplicaciones Tecnol\'{o}gicas y Desarrollo Nuclear (CEADEN), Cubaenerg\'{\i}a, Cuba;
Ministry of Education, Youth and Sports of the Czech Republic, Czech Republic;
The Danish Council for Independent Research | Natural Sciences, the VILLUM FONDEN and Danish National Research Foundation (DNRF), Denmark;
Helsinki Institute of Physics (HIP), Finland;
Commissariat \`{a} l'Energie Atomique (CEA) and Institut National de Physique Nucl\'{e}aire et de Physique des Particules (IN2P3) and Centre National de la Recherche Scientifique (CNRS), France;
Bundesministerium f\"{u}r Bildung und Forschung (BMBF) and GSI Helmholtzzentrum f\"{u}r Schwerionenforschung GmbH, Germany;
General Secretariat for Research and Technology, Ministry of Education, Research and Religions, Greece;
National Research, Development and Innovation Office, Hungary;
Department of Atomic Energy Government of India (DAE), Department of Science and Technology, Government of India (DST), University Grants Commission, Government of India (UGC) and Council of Scientific and Industrial Research (CSIR), India;
Indonesian Institute of Science, Indonesia;
Istituto Nazionale di Fisica Nucleare (INFN), Italy;
Institute for Innovative Science and Technology , Nagasaki Institute of Applied Science (IIST), Japanese Ministry of Education, Culture, Sports, Science and Technology (MEXT) and Japan Society for the Promotion of Science (JSPS) KAKENHI, Japan;
Consejo Nacional de Ciencia (CONACYT) y Tecnolog\'{i}a, through Fondo de Cooperaci\'{o}n Internacional en Ciencia y Tecnolog\'{i}a (FONCICYT) and Direcci\'{o}n General de Asuntos del Personal Academico (DGAPA), Mexico;
Nederlandse Organisatie voor Wetenschappelijk Onderzoek (NWO), Netherlands;
The Research Council of Norway, Norway;
Commission on Science and Technology for Sustainable Development in the South (COMSATS), Pakistan;
Pontificia Universidad Cat\'{o}lica del Per\'{u}, Peru;
Ministry of Science and Higher Education, National Science Centre and WUT ID-UB, Poland;
Korea Institute of Science and Technology Information and National Research Foundation of Korea (NRF), Republic of Korea;
Ministry of Education and Scientific Research, Institute of Atomic Physics and Ministry of Research and Innovation and Institute of Atomic Physics, Romania;
Joint Institute for Nuclear Research (JINR), Ministry of Education and Science of the Russian Federation, National Research Centre Kurchatov Institute, Russian Science Foundation and Russian Foundation for Basic Research, Russia;
Ministry of Education, Science, Research and Sport of the Slovak Republic, Slovakia;
National Research Foundation of South Africa, South Africa;
Swedish Research Council (VR) and Knut \& Alice Wallenberg Foundation (KAW), Sweden;
European Organization for Nuclear Research, Switzerland;
Suranaree University of Technology (SUT), National Science and Technology Development Agency (NSDTA) and Office of the Higher Education Commission under NRU project of Thailand, Thailand;
Turkish Atomic Energy Agency (TAEK), Turkey;
National Academy of  Sciences of Ukraine, Ukraine;
Science and Technology Facilities Council (STFC), United Kingdom;
National Science Foundation of the United States of America (NSF) and United States Department of Energy, Office of Nuclear Physics (DOE NP), United States of America.

%% file: Alice_Authorlist_2020-12-27.tex
\begin{flushleft}

S.~Acharya$^{\rm 142}$, 
D.~Adamov\'{a}$^{\rm 97}$, 
A.~Adler$^{\rm 75}$, 
J.~Adolfsson$^{\rm 82}$, 
G.~Aglieri Rinella$^{\rm 35}$, 
M.~Agnello$^{\rm 31}$, 
N.~Agrawal$^{\rm 55}$, 
Z.~Ahammed$^{\rm 142}$, 
S.~Ahmad$^{\rm 16}$, 
S.U.~Ahn$^{\rm 77}$
% ...

\bigskip

\bigskip 

\textbf{\Large Affiliation Notes}

\bigskip 

% ...
$^{\rm V}$ Also at: Institute of Theoretical Physics, University of Wroclaw, Poland\\

\bigskip

\bigskip 

\textbf{\Large Collaboration Institutes}

\bigskip 

% ...
$^{16}$ Department of Physics, Aligarh Muslim University, Aligarh, India\\
% ...
$^{31}$ Dipartimento DISAT del Politecnico and Sezione INFN, Turin, Italy\\
% ...
$^{35}$ European Organization for Nuclear Research (CERN), Geneva, Switzerland\\
% ...
$^{55}$ INFN, Sezione di Bologna, Bologna, Italy\\
% ...
$^{75}$ Johann-Wolfgang-Goethe Universit\"{a}t Frankfurt Institut f\"{u}r Informatik, Fachbereich Informatik und Mathematik, Frankfurt, Germany\\
% ...
$^{77}$ Korea Institute of Science and Technology Information, Daejeon, Republic of Korea\\
% ...
$^{82}$ Lund University Department of Physics, Division of Particle Physics, Lund, Sweden\\
% ...
$^{97}$ Nuclear Physics Institute of the Czech Academy of Sciences, \v{R}e\v{z} u Prahy, Czech Republic\\
% ...
$^{142}$ Variable Energy Cyclotron Centre, Homi Bhabha National Institute, Kolkata, India\\
% ...

\end{flushleft} 